# 360° domain wall generation in the soft layer of magnetic tunnel junctions

M. Hehn, D. Lacour, F. Montaigne, J. Briones

*Laboratoire de Physique des Matériaux (LPM), Nancy-Université, CNRS, Boulevard des Aiguillettes, B.P. 239, F-54506 Vandœuvre lès Nancy, France*

R. Belkhou, S. El Moussaoui, F. Maccherozzi

*Synchrotron SOLEIL, L'Orme des Merisiers Saint-Aubin, 91192 Gif-sur-Yvette, France and, Synchrotron ELETTRA, AREA Science Park, 34012 Basovizza, Trieste, Italy*

N. Rougemaille

*Institut Néel, CNRS & Université Joseph Fourier, 25 rue des Martyrs, BP 166, F-38042 Grenoble cedex 9, France*

High spatial resolution X-ray photo-emission electron microscopy technique has been used to study the influence of the dipolar coupling taking place between the NiFe and the Co ferromagnetic electrodes of micron sized, elliptical shaped magnetic tunnel junctions. The chemical selectivity of this technique allows to observe independently the magnetic domain structure in each ferromagnetic electrode. The combination of this powerful imaging technique with micromagnetic simulations allows to evidence that a 360° domain wall can be stabilized in the NiFe soft layer. In this letter, we discuss the origin and the formation conditions of those 360° domain walls evidenced experimentally and numerically.

Most studies concerning the micromagnetic properties of nanostructures are performed on single ferromagnetic layers [1, 2]. However, spintronics applications generally require structures with several magnetic layers, separated by thin, non magnetic spacers. In these structures, magnetic couplings may appear between the ferromagnetic layers. This is emphasised in nanostructures having in-plan magnetizations for which an antiferromagnetic dipolar coupling, originating from the stray fields at the nanostructure edges, is no more to be neglected. As a consequence, antiparallel alignment of the two layers is observed in micron [3] and submicron tunnel junctions and spin valves [4]. The domain structure induced by this coupling has been studied using various techniques such as Lorentz microscopy [3, 5, 6], electron holography [4] or magnetic force microscopy [7]. But none of them is able to give unambiguously the configuration in each magnetic layer. In this letter, we used the chemical selectivity and the high spatial resolution of the X-ray photo-emission electron microscopy technique (XMCD-PEEM) to study the influence of the dipolar coupling taking place between a FeNi (Py: soft layer) and a Co (hard layer) layer when inserted in a elliptical shaped magnetic tunnel junction (MTJ). The comparison of our experimental results to micromagnetic calculations shed light on the previous observations of 360° domain walls (DW) [5, 6], on their formation conditions and on their statistics of occurrence.

The MTJ multilayer stack grown in an UHV Alliance Concept sputtering plan on a Si (100) substrate is composed of a Ta(5) /Al$_2$O$_3$(2) /Co(4) /Al$_2$O$_3$(2) /Fe$_{20}$Ni$_{80}$(4) / Ru(2) multilayer stack (layer thicknesses in nm). The coercive fields of the continuous Co and Py layers before etching into an elliptical shape are respectively 50 Oe and 20 Oe. The ferromagnetic coupling field between the two layers is less then 5 Oe. 1x3 µm² ellipses have been patterned by electron beam lithography using a JEOL 6500F SEM. These structures are defined using an Al mask and a subsequent Ar ion beam etching down to the substrate. The ferromagnetic and the Ru capping layer thicknesses have been chosen in order to optimize the XMCD-PEEM contrasts in both ferromagnetic electrodes.

To determine the onset of the spin structure within each layer of the MTJ, the samples were imaged using X-ray Magnetic Circular Dichroism Photoemission Microscopy (XMCD-PEEM). The photon energy was set to the Ni L$_3$ and Co L$_3$ edges to image selectively the domain structure of respectively the Py and the Co layers. The second electron yield in this microscopy technique is proportional to the scalar product of the magnetisation and the helicity of the elliptically polarised synchrotron light ($\sigma.M$). The yield difference between opposite helicities is visible as a magnetic contrast in the images [8].

Figure 1 shows XMCD-PEEM images taken at remnance recorded at the Co and Ni edges after saturation with a 1 kOe field applied parallel to the ellipses long axis. Here, the X-ray beam direction is aligned with the ellipses long axis. So the observed contrast is due to the magnetization component parallel to the long axis. In the following and for all the experiments, the Co magnetization is uniform and oriented along the saturating field (white contrast in Fig 1 a). Regarding the Py layer, some elements show a single domain state but magnetized in the opposite direction. Other ellipses present an inhomogeneous magnetization configuration of the soft Py layer. More insight in this domain structure is gained after a 90° rotation of the sample with respect to the X-ray beam direction. Then, the magnetic contrast arises mostly from the transverse component of the magnetisation (along the ellipses short axis). Figures 2 a. and 2 b. show XMCD-PEEM images recorded at the Ni edge for two ellipses exhibiting a non-uniform state. Black/white or white/black contrasts in the Py layer are signatures of clockwise (CW) and counter clockwise (CCW) 360° domain walls as sketched in fig 2. d.

This existence of 360° DWs is related to the topological impossibility to unwind such kind of structure. Understanding the presence or the absence of these 360° DWs requires the study of the Py magnetization reversal dynamics. Therefore, we have modeled our layered ellipses using 3D micromagnetic simulations based on the OOMMF package [9]. *In silico* experiences show that the highly inhomogeneous dipolar field originating from the Co layer edges activates the reversal process of the soft layer. The Py magnetization reversal initiates at the ellipses extremities where this field is maximum. When the saturating field is switched off, two domains nucleate at each extremity and propagate towards the core. The chirality of the two generated 180° DWs is determined by the initial curling direction of magnetization at the ellipse extremities. If both walls have the same chirality when joining the ellipses centre, they collapse leading to a single domain state. However if the two walls have opposited chiralities they will join forming a 360° DW. The initial domains are then shrinked to the DW core. The nucleation of the two domains by opposite curling direction is thus a mandatory condition to the formation of a 360° DW. Figure 2 c. shows such a stable domain wall obtained with the 3D OOMMF code. The intensity profiles across the two kinds of DWs (CW and CCW) observed by XMCD-PEEM are plotted in fig. 2 e. For comparison, the profile of the CCW DW obtained numerically is also plotted in fig. 2 e. A good agreement is observed between the experimental and theoretical 360° DW lateral extension.

As mentioned earlier, all the ellipses do not exhibit the same magnetic state at zero applied field: some have a single domain Py layer while others host a 360° DW. Figure 3 shows two pictures of the same 4 ellipses recorded in the 90° configuration in order to highlight only the structure of the DWs . The 4 ellipses, chosen for their behavior diversity, have been saturated twice along the long ellipse axis. For the γ and δ ellipses, the same magnetic configuration is obtained, respectively a single domain state and a 360° DW. For the α ellipse, a DW with the same chirality is observed but its position has changed after the second magnetic saturation. Finally, for the β ellipse, a single domain is first observed while a DW is obtained after the second saturation. The dynamical process leading to the formation of a DW is thus partly stochastic.

The statistics of DW formation in the soft layer is then studied by applying 3 saturation/measurement cycles on a set of 65 objects. We observed that 74% of the ellipses systematically adopt exactly the same configuration (γ and δ like), 8% have always a DW but located at different positions (α like) and 18% are either in a single domain state or host a DW (β like). Moreover, for a given ellipse, no change of DW chirality is observed.

The γ and δ like ellipses, exhibiting the same configurations, are composed of three subgroups: 50% of the elements have a single domain state, 27% stabilize a CW DW, and 23% a CCW DW. As the presence of a DW is linked to the initial direction of curling, these statistics prove that for a majority of elements, the curling direction is determined uniquely for each ellipses extremity. Furthermore it appears that each curling direction is equally probable on the observed array. Indeed an equal probability for each curling direction implies an occurrence probability of 0.5 for the single domain state and 0.25 for each DW chirality. This corresponds within the statistical errors to the population measured experimentally.

This result is in contradiction with the micromagnetic simulations. For perfectly symmetric elements, the curling direction is determined uniquely by the torque exerted by the dipolar field originating from the Co layer. Since this field has opposed vertical components at the two extremities of the ellipses, the magnetization curling directions are also opposed and only CCW DW should be stabilized. This is obviously not the case experimentally. Therefore some other factors, more related to the local structure of the ellipses and to their topology, may influence the curling direction : (i) micromagnetic simulations show that shape defects as small as 40 nm induced by the lithographic process and located at the ellipse ends, can overcome the intrinsic influence of the torque; (ii) small local magnetic anisotropy variation within the ellipse due to the polycrystalline nature of the Py layer as proved by the observed ripple structure (see weak contrast variations on fig. 2 and 3). These sources of anisotropy can overcome the intrinsic influence of the Co torque and each ellipse extremity may thus have a specific anisotropy direction. Those local sources of anisotropy (shape and/or magnetocristalline) break the symmetry of the system. For 74% of the elements exhibiting always the same magnetic configuration, the initial curling direction at each ellipse extremity determines the zero field state. In this case, the observed magnetic configurations are induced by the local magnetic anisotropy at the ellipses extremities.

For 8% of the ellipses ($\alpha$ case), the position of the DW changes for different saturations. This highlights the importance of the propagation process. Indeed, the propagation of DW is significantly affected by defects and particularly in reduced dimension structures [10]. Due to the random distribution of defects, each 180° DW propagating from each extremity experiences a specific "pinning potential". The final position of the 360° DW is the intercept of the two 180° DW trajectories. It thus varies from ellipse to ellipse. As the thermal activation plays an important role in the depinning process, the trajectory and the final position of 360° DWs have a stochastic nature. Finally, a significant number of ellipses ($\beta$ case, 18% of ellipses) presents either a 360° DW or a single domain state. In this case, the curling direction is not uniquely determined by the local magnetic anisotropy at one ellipses extremity: the magnetization curling direction could be driven by a stochastic effect as thermal fluctuations.

In conclusion, we have presented systematic studies of ellipse shaped magnetic tunnel junctions. We show that the formation of 360° DWs in the soft layer is not only linked to the torque exerted from the hard magnetic layer as expected from micromagnetic simulations. The presence of DWs is mainly influenced by the local magnetic anisotropies existing at each extremity of the elements. Controlling the nucleation/propagation of such DWs is essential for magneto-transport measurements or for devices based on the switching of the two layers of a magnetic tunnel junction [11]. The authors would like to thank J. Vogel for fruitful discussions and G. Lengaigne for technical assistance. The XMCD-PEEM measurements have been performed on the SOLEIL's X-PEEM microscope installed on the Nanospectroscopy beamline at the synchrotron Elettra - Italy.

Figure captions

Figure 1:

XMCD-PEEM images of 4 ellipses recorded at the Co (a) and at the Ni (b) edges. The photons incidence direction is aligned along the ellipses long axis from the top left corner. The white and black contrasts correspond to magnetization components aligned along the long ellipses axis, parallel or antiparallel to the saturating field.

Figure 2:

a) and b) XMCD-PEEM images of 2 ellipses (1µm x 3 µm) measured at the Ni edge. The gray level distribution corresponds to the scalar projection of the local magnetization with respect to the light incidence. c) OOMMF simulation. The grey scale corresponds to the magnetization component along the short axis direction. d) Sketches representing the two possible 360° DWs either CW or CCW. e) Experimental and simulated intensity profiles across a 360° DW.

Figure 3 :

XMCD-PEEM images of the same ellipses recorded after the two cycles of identical field history. The ellipses have been images with the incoming light aligned along the short ellipse axis in order to highlight the magnetic structure of the DWs.

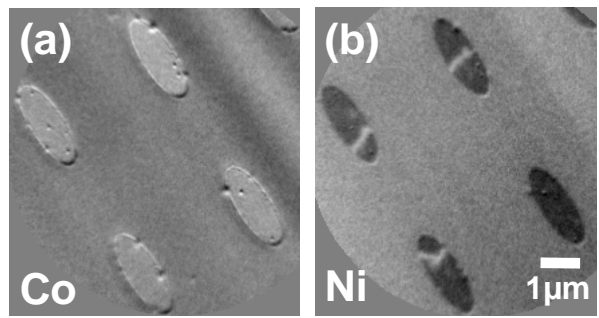

Figure 1, M. Hehn et al.

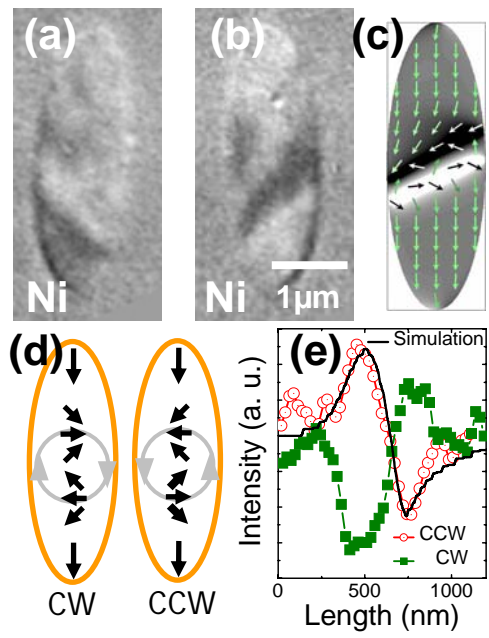

Figure 2, M. Hehn et al.

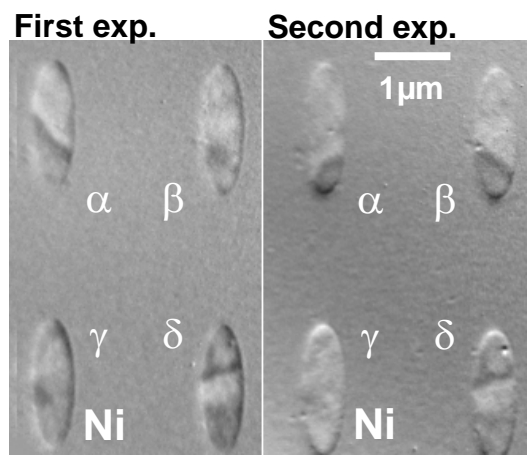

Figure 3, M. Hehn et al.